\documentstyle{mn}
\input{epsf}
\def\ffrac#1#2{{\textstyle\frac{#1}{#2}}}
\def\la{\mathrel{\mathpalette\fun <}}
\def\ga{\mathrel{\mathpalette\fun >}}
\def\fun#1#2{\lower3.6pt\vbox{\baselineskip0pt\lineskip.9pt
\ialign{$\mathsurround=0pt#1\hfill##\hfil$\crcr#2\crcr\sim\crcr}}}
\def\APM {{\rm APM}}
\def\CDM {{\rm CDM}}
\def\dof {{\rm d.o.f.}}
\def\en {{\rm e}}
\def\H {{\rm H}}
\def\LSS {{\rm LSS}}
\def\lin {{\rm lin}}

\def\mpc {h^{-1} {\rm Mpc}}

\def\impc {h\,{\rm Mpc}^{-1}}
\def\rmB {{\rm B}}
\def\rmc {{\rm c}}
\def\rmd {{\rm d}}
\def\rme {{\rm e}}
\def\rmm {{\rm m}}
\def\sta {{\rm s}}
\begin{document}
\journal{OUTP-00-33-P}
\title[On the APM power spectrum and the CMB anisotropy]
      {On the APM power spectrum and the CMB anisotropy: 
       Evidence for a phase transition during inflation?} 
\author[J. Barriga, E. Gazta\~naga, M.G. Santos and S. Sarkar]
 {J. Barriga$^{1,2}$, E. Gazta\~naga$^{1,2}$, M.G. Santos$^3$ and 
  S. Sarkar$^3$ \\
$^{1}$ Institut d'Estudis Espacials de Catalunya, IEEC/CSIC, 
 Edf. Nexus-201 - c/ Gran Capitan 2-4, 08034 Barcelona, Spain \\
$^{2}$ INAOE, Astrofisica, Tonantzintla, Apdo Postal 216 y 51, 
 Puebla 7200, Mexico \\
$^{3}$ Theoretical Physics, Department of Physics, University of Oxford, 
 1 Keble Road, OX1 3NP, UK}
\date{\today}
\pubyear{2000}
\maketitle
\begin{abstract}
 Adams et al. (1997b) have noted that according to our current
 understanding of the unification of fundamental interactions, there
 should have been phase transitions associated with spontaneous
 symmetry breaking {\em during} the inflationary era. This may have
 resulted in the breaking of scale-invariance of the primordial
 density perturbation for brief periods. A possible such feature was
 identified in the power spectrum of galaxy clustering in the APM
 survey at the scale $k\sim0.1\,\impc$ and it was shown that the
 secondary acoustic peaks in the power spectrum of the CMB anisotropy
 should consequently be suppressed. We demonstrate that this
 prediction is confirmed by the recent Boomerang and Maxima
 observations, which favour a step-like spectral feature in the range
 $k\sim(0.06-0.6)\impc$, independently of the similar previous
 indication from the APM data. Such a spectral break enables an
 excellent fit to both APM and CMB data with a baryon density
 consistent with the BBN value. It also allows the possibility of a
 matter-dominated universe with zero cosmological constant, which we
 show can now account for even the evolution of the abundance of rich
 clusters.
\end{abstract}

\section{Introduction}

It is commonly assumed by astronomers that inflation predicts a
scale-invariant `Harrison-Zeldovich' (H-Z) spectrum of scalar density
perturbations, $\delta_\H^2(k)\propto\,k^{n-1}$ with $n=1$; for
example this is a standard input in calculations of the expected
large-scale structure (LSS) and cosmic microwave background (CMB)
anisotropy. In fact in the usual `slow-roll' inflationary scenario
there is a gradual steepening of the spectrum with increasing $k$
(decreasing scale), as the end of inflation is approached
\cite{mc81,h82,s82,gp82,bst83}. This cannot be adequately modelled by
a spectrum with constant ``tilt'' ($n<1$) as is occasionally adopted,
since the steepening is {\em scale-dependent} for any polynomial
inflaton potential.\footnote{The index $n$ is scale-independent only
for an exponential potential (`power-law inflation'); moreover $n$ can
be very close to unity if inflation ends not through the steepening of
the inflaton potential but, for example, due to the dynamics of a
second scalar field (`hybrid inflation'). The scale-dependence of $n$
in various inflationary models has been reviewed by Lyth \& Riotto
(1999).}  Even such small departures from scale-invariance can be
quite significant, for example in a supergravity model with a cubic
inflaton potential, Adams, Ross \& Sarkar (1997a) found
$n\simeq(N-2)/(N+2)$, where $N\sim51+\ln(k^{-1}/3000\,\mpc)$ is the
number of e-folds of expansion from the end of inflation. This gradual
suppression of small-scale power was shown to be adequate to reconcile
the COBE-normalised standard cold dark matter (sCDM) model with LSS
data, in particular the power spectrum of galaxy clustering, $P(k)$,
the abundance of rich clusters quantified by the parameter $\sigma_8$,
and large-scale streaming velocities, $\sigma_v(40\,\mpc)$. We emphasise
that this model is however ruled out by the same data if a
scale-invariant spectrum is assumed \cite{ebw92}. Such conclusions are
clearly not robust given that we have as yet no standard model of
inflation.

Subsequently it was realized \cite{ars97b} that the spectrum may not
be even {\em scale-free} because the rapid cooling of the universe
during primordial inflation can result in spontaneous symmetry
breaking phase transitions which may interrupt the slow roll of the
inflaton field for brief periods. This is in fact inevitable in models
based on $N=1$ supergravity, the phenomenologically successful
extension of the Standard Model of particle physics and the effective
field theory below the Planck scale. During inflation, the large
vacuum energy breaks supersymmetry giving otherwise massless fields
(`flat directions') a mass of order the Hubble parameter
\cite{dfn84,cfkrs84}, causing them to evolve rapidly to the asymmetric
global minima of their scalar potential. Such `intermediate scale'
fields are generic in models derived from superstring/M-theory and
have gauge and Yukawa couplings to the thermal plasma so are initially
confined at the symmetric maxima of their scalar
potentials. Consequently it takes a (calculable) finite amount of
cooling before the thermal barrier disappears and they are free to
evolve to their minima \cite{y86,bclp96}. When a symmetry breaking
transition occurs, the mass of the inflaton field changes suddenly
(through couplings in the K\"ahler potential), temporarily violating
the slow-roll conditions and interrupting inflation (Adams et
al. 1997b).\footnote{When (re)heating occurs at the end of inflation
such fields may again be forced back to the symmetric maximum,
undergoing symmetry breaking a second time when the universe cools
down to the electroweak scale in the radiation-dominated era and
driving a late phase of `thermal inflation' \cite{ls96}.} Thus the
density perturbation is expected to have a (near) H-Z spectrum for the
first $\sim10$ e-folds of expansion followed by one or more sudden
departures from scale-invariance lasting $\sim1$ e-fold. In order for
such spectral features to be observable in the LSS or CMB, it is of
course necessary that they occur within the last $\sim50$ e-folds of
inflation, corresponding to spatial scales going up to the present
Hubble radius $H_0^{-1}\sim3000\,\mpc$. Since the density perturbation
can be observed on scales from the Hubble radius down to $\sim1\,\mpc$,
corresponding to about 8 e-folds of expansion, it would be not
unreasonable to expect at least one such spectral break to be seen
today.

Motivated by this an attempt was made to recover the primordial
perturbation spectrum from extant observations of LSS and CMB
anisotropy, specifically the APM survey \cite{apm90a} and the COBE
observations \cite{cobe92}. We recall that the spectrum of rms mass
fluctuations at the present epoch (per unit logarithmic interval of
wavenumber $k$) is
\begin{equation}
 \Delta^2 (k) \equiv \frac{k^3 P(k)}{2\pi^2} 
           = \delta^2_\H (k)\ T^2 (k) \left(\frac{k}{H_0}\right)^{3+n}\ ,
\label{deltak}
\end{equation}
where the density perturbation is evaluated at the present Hubble
radius, i.e. at $k=H_0$. The (dimensionless) matter `transfer
function' for CDM models can be approximated by \cite{be84}
\begin{equation}
T_{\CDM}(k) \simeq \left[1 + \left\{a k + (b k)^{3/2} + (c k)^2
\right\}^{\nu}\right]^{-1/\nu}\ ,
\label{Tk}
\end{equation}
with $a=6.4\Gamma^{-1}\,\mpc$, $b=3\Gamma^{-1}\,\mpc$,
$c=1.7\Gamma^{-1}\,\mpc$ and $\nu=1.13$. Here the `shape parameter' is
defined as \cite{bj99} 
\begin{equation}
\Gamma = \Omega_\rmm{h}\rme^{-[\Omega_\rmB(1+\sqrt{2h}/\Omega_\rmm)-0.06]},
\label{Gamma}
\end{equation}
where $h\equiv\,H_0/100\,{\rm Km\,s}^{-1}{\rm Mpc}^{-1}$ is the Hubble
parameter and $\Omega_\rmm=\Omega_\CDM+\Omega_\rmB$, $\Omega_\CDM$
($\Omega_\rmB$) being the fraction of the critical density in cold
(baryonic) matter. Thus $\Gamma=0.5$ for sCDM which assumes
$\Omega_\rmB=0.03$, $\Omega_\rmm=1$ and $h=0.5$. 

Following Baugh \& Gazta\~naga (1996), the three-dimensional
$P_{\APM}(k)$ inferred from the angular correlation function of
galaxies in the APM survey \cite{be93} was subjected to an inversion
procedure \cite{jmw95,pd96} to recover the linear spectrum of
fluctuations and obtain its spectral index,
$\rmd\ln{P(k)}/\rmd\ln{k}$, as a function of $k$. Assuming that any
bias between APM galaxies and dark matter is {\em scale-independent},
the primordial spectral index $n(k)$ was then obtained simply by
subtracting off the slope of the transfer function (\ref{Tk}) for each
value of $k$ (Adams et al. 1997b).

The key finding was that $n_{\APM}(k)$ shows a dramatic departure from
the H-Z value of $n=1$ in the range $k\sim(0.05-0.6)\impc$, dropping
to a value around zero ($\sim3\sigma$ below unity) at
$k\sim0.1\,\impc$. The sharp drop occurs over $\sim1$ e-fold of
expansion as is indeed expected in the phase transition model. This
feature in $P_{\APM}(k)$ had been noted by Baugh \& Efstathiou (1993)
and was also identified by Peacock (1997) in the power spectrum of
IRAS galaxies after correcting for redshift-space distortions. Given
that the feature appears on the scale of the Hubble radius at the
epoch of (dark) matter domination, it is perhaps natural to interpret
it as reflecting a departure from the sCDM paradigm as these authors
did. However Adams et al. (1997b) pursued the alternative possibility
that it is in fact present in the primordial spectrum and asked what
would be the expectation for the angular anisotropy of the CMB if this
were so. Using the COSMICS code \cite{b95} they calculated its angular
power spectrum and found that the height of the secondary acoustic
peaks is suppressed by a factor of $\sim2$ relative to the prediction
of the standard COBE-normalised sCDM model which assumes a scale-free
H-Z spectrum.

We reexamine this prediction in the light of recent observations by
the Boomerang \cite{boom00} and MAXIMA-1 \cite{max00} experiments
which {\em do} indicate a significant suppression of the second
acoustic peak. It has been noted by many authors
\cite{tz00,boomest01,maxest00,hfzt00,boomaxest00,emmmp01,ps00} that
this requires a high baryon density, significantly above the value
required by light element abundances following from standard Big Bang
nucleosynthesis (BBN). As the latter is widely considered to be one of
the ``pillars'' of modern cosmology, it is important to ask whether
this conflict may not be resolved by relaxing the assumption made in
these analyses that the primordial spectrum of perturbations is
scale-free.

\section{Reconstructing the primordial spectrum}

We begin by describing the range of values that we consider for the
cosmological parameters. This represents the set of priors, resulting
from other observations, that we adopt for the reconstruction of the
primordial spectrum. In \S2.2, we discuss how LSS data can be used to
recover the primordial spectrum and predict the CMB anisotropy
(\S2.3). In \S2.4 we present a simple parameterization for the
primordial spectrum and use this to fit the new CMB as well as APM
data. Finally in \S2.5 we confront the observed abundance of rich
clusters with the expectations in our model.

\subsection{Cosmological parameters}

Let us first briefly discuss the choice of cosmological parameters. As
mentioned before, $\Gamma=0.5$ in the sCDM model, but
$\Gamma\simeq0.2$ gives in fact a better fit to the APM data
\cite{apm90b,ebw92,be93,pd94} and can be realized e.g. in a low
density model with $\Omega_\rmm\sim0.4$.\footnote{However structure
formation does not necessarily require a low density universe as is
often claimed (e.g. Bahcall et al. 1999); for example with a gradually
tilting primordial spectrum obtained from `new' supergravity
inflation, the small-scale power is sufficiently suppressed for a
$\Omega_\rmm=1$ CDM model to be compatible with LSS data (Adams et
al. 1997a, see also White et al. 1995). Such `tilt' can preferentially
suppress the second acoustic peak if the baryon fraction is high
(Adams et al. 1997a, see also White et al. 1996). Several groups have
studied the implications of tilt for the recent CMB data (e.g. Kinney,
Melchiorri \& Riotto 2000, Covi \& Lyth 2000, Tegmark, Zaldariagga \&
Hamilton 2001).} The alternative possibility of a Hubble parameter as
low as $h\sim0.3$ \cite{bbst95} is definitively ruled out by the HST
Key Project data on 25 galaxies which indicates $h=0.71\pm0.06$
\cite{hst00}. Weighting such direct observations against other
fundamental physics approaches, Primack (2000) advocates
$h=0.65\pm0.08$. However a value as low as 0.45 \cite{psst00} or as
high as 0.9 \cite{tbad00} still seems quite possible so the quoted
uncertainty cannot be interpreted as a gaussian standard
deviation. Instead we assume that all values in the following
``$2\sigma$'' range are equally probable:
\begin{equation}
 h = 0.49 - 0.81 .
\label{h}
\end{equation}
The current preference is in fact for a cosmological constant
dominated flat universe with $\Omega_\Lambda \simeq2/3$ and
$\Omega_\rmm\simeq1/3$ (e.g. Bahcall et al. 1999). The Boomerang and
MAXIMA observations of the first acoustic peak indicate that the
universe is indeed flat \cite{boomaxest00}, but we will consider other
$\Lambda$CDM models with values of $\Omega_\Lambda$ ranging down to
zero, maintaining $\Omega\rmm+\Omega_\Lambda=1$. This is motivated by
our concern that the only {\em direct} evidence suggesting
$\Omega_\Lambda\sim0.7$ is the small curvature of the Hubble diagram
for Type Ia supernovae at redshift $z\sim1$ \cite{scp99,hzs98} which
may be subject to unrecognised systematic effects. (We note that other
arguments for a low $\Omega_\rmm$ from the evolution of the cluster
number density \cite{ecfh98} or the clustering in the Lyman-$\alpha$
forest \cite{wchkp99} rest on the {\em assumption} of a scale-free
spectrum.) Of course for any particular choice of
$(\Omega_\rmm,\Omega_\Lambda)$, one has to check that the age of the
universe exceeds the inferred age of globular clusters, $t_{\rm
GC}=12.8\pm1$~Gyr \cite{k00}.

Another important parameter for the present study is the BBN value of
$\Omega_\rmB$. This has undergone substantial revision since the first
studies of sCDM so we briefly review the present situation. The baryon
density which provides a good fit to the {\em inferred} primordial
abundances of D, $^4$He and $^7$Li in standard BBN is \cite{flsv98}
\begin{equation}
 \Omega_\rmB = (0.019^{+0.0013}_{-0.0012})h^{-2},
\label{omegab}
\end{equation}
taking all (correlated) reaction rate uncertainties into
account.\footnote{Subsequently Burles et al. (1999) suggested that
some of the reaction rate uncertainties had been overestimated; thus
they inferred a somewhat tighter range
$\Omega_\rmB=(0.019\pm0.0009)h^{-2}$. Izotov et al. (2000) found
however a better fit with
$\Omega_\rmB=(0.016^{+0.0018}_{-0.0016})h^{-2}$, taking
Y($^4$He)=$0.2452\pm0.0015$ from observations of the two most
metal-deficient BCGs, D/H=$(4.35\pm0.43)\times10^{-5}$ from the
analysis of QAS data using kinematic models incorporating
`mesoturbulence' \cite{lkt99}, and
$^7$Li/H=$(2.24\pm0.57)\times10^{-10}$ allowing for some depletion due
to rotation in Pop II stars \cite{vc98}.} This fit is based on the
`low' value of D/H=$(3.4\pm0.25)\times10^{-5}$ found \cite{bt98} in
two quasar absorption systems (QAS) at high redshift [and subsequently
in a third one \cite{ktblo00}], the `high' value of the mass fraction
Y($^4$He)=$0.243\pm0.003$ inferred \cite{itl98} from observations of
metal-poor blue compact galaxies (BCGs), and the `Spite plateau' of
$^7$Li/H=$(1.73\pm0.12)\times10^{-10}$ seen in Pop~II stars
\cite{bm97}. Note however that an improved fit is obtained
\cite{flsv98} at a significantly smaller value of
$\Omega_\rmB=(0.0065^{+0.0010}_{-0.00064})h^{-2}$ if one adopts the
`high' value of D/H=$(1.9\pm0.4)\times10^{-4}$ claimed \cite{schr94}
in another QAS [but disputed subsequently \cite{bkt99}, although
consistent with a possible high value \cite{tblfws99} in yet another
QAS], and the `low' value of Y($^4$He)=$0.234\pm0.005$ obtained
\cite{oss97} from analysis of older BCG data [and supported by another
recent observation \cite{ppr00}]. An independent lower limit of
$\Omega_\rmB>0.0125h^{-2}$\cite{wmhk97} set by observations of the
Lyman-$\alpha$ forest favours the higher value in
Equation~(\ref{omegab}), which also allows a better fit to the
observed distribution of line widths in comparison with model
hydrodynamic simulations (e.g. Theuns et al. 1999). An audit of
luminous matter in the universe \cite{fhp98} gives a much weaker lower
limit of $\Omega_\rmB\ga0.007$. A conservative upper limit on the
baryon density is \cite{ks96}
\begin{equation}
 \Omega_\rmB < 0.033\,h^{-2},
\label{omegabmax}
\end{equation}
if the primordial deuterium abundance is bounded from below by its
typical interstellar value D/H=$(1.5\pm0.2)\times10^{-5}$
\cite{m92}. Therefore we will consider a wide range of possible values
in our analysis, but adopt $\Omega_\rmB=0.045$, $h=0.65$ as the
default values (so that $\Omega_\CDM=\Omega_\rmm-0.045$) when not
stated otherwise.

\subsection{Primordial spectrum from the APM data}

There are several arguments (reviewed in Appendix~A) that on scales
$0.01\,\impc\la\,k\la\,0.6\,\impc$, which are at most weakly non-linear,
the APM galaxy power spectrum $P_\APM(k)$ \cite{be93} is an unbiased
(or moderately linearly biased) tracer of the mass. The linear power
spectrum recovered under this assumption from $P_\APM(k)$, using the
method of Jain et al. (1995), is well fitted in this range by
\cite{bg96}:
\begin{equation}
 P_{\lin}(k) \simeq \frac{7\times10^{5} k\,(\,\mpc)^3}
                    {\left[1+(k/k_\rmc)^2\right]^{1.6}}\ ,
\label{pklin}
\end{equation}
where $k_\rmc=150(H_0/c)\simeq0.05\,\impc$. We will use the common
convention of expressing the matter power spectrum as the primordial
spectrum of matter fluctuations $P^0(k)\propto\,k\delta^2_\H(k)$,
folded with the transfer function (\ref{Tk}):
\begin{equation}
 P_\lin(k) \equiv P^0(k)\,T^2(k) = A k^n T^2 (k), 
\end{equation}
where the (dimensionful) normalisation constant $A$ is determined by
the large-scale CMB anisotropy. [For sCDM with $n=1$ normalised to the
COBE 4-year data \cite{cobe96}, $\delta_\H=1.92\times10^{-5}$ with 9\%
uncertainty \cite{bw97} so $A=(5.9\pm1.1)\times10^5(\,\mpc)^4$.] 

Given some estimate of the (linear) matter power spectrum, as in
Equation~(\ref{pklin}), the standard approach is to assume a
scale-invariant initial spectrum, i.e. $ P^0(k) \propto\,k$, and
extract the value of $\Gamma$ that best fits the resulting constraint
on $T(k)$. Alternatively one can extract $P^0(k) (=P_{\lin}/T^2$) for
a specific choice of $\Gamma$, i.e. for a given transfer function. We
apply this latter approach to the APM data by simply taking the ratio
of Equation~(\ref{pklin}) to Equation~(\ref{Tk}), obtaining:
\begin{eqnarray}
 P^0(k) = \left\{ \begin{array}{ll}
  A_1 k, & \mbox{$k<k_1$,}\\
  A k \frac{\left[1+(ak+(bk)^{3/2}+(ck)^2)^u\right]^{2/u}}
   {\left[1+(k/k_\rmc)^2\right]^{1.6}}, & \mbox{$k_1 \leq k \leq k_2$,}\\
  A_2 k, & \mbox{$k_2<k$,}
\end{array} \right.
\label{pkprim}
\end{eqnarray}
where $A=7\times10^{5} (\,\mpc)^3$ and we have adopted a suitably
normalised H-Z spectrum outside the range ($k<k_1=0.01\,\impc$,
$k>k_2=0.6\,\impc$) accessible to galaxy clustering observations.

\begin{figure}
\epsfxsize\hsize\epsffile{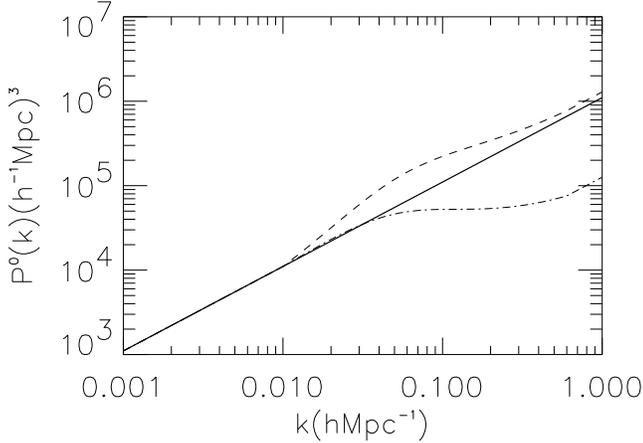}
\caption[junk]{Reconstruction of the primordial density power spectrum
 from the APM data, adopting a CDM shape parameter $\Gamma$ of 0.5
 (dot-dashed line) and 0.2 (dashed line). A Harrison-Zeldovich
 spectrum (full line) is shown for comparison.}
\label{fig1}
\end{figure}

Figure~\ref{fig1} shows the recovered spectrum (\ref{pkprim}) for two
choices of $\Gamma$ corresponding to the sCDM model and a low density
variant. Note that the $\Gamma=0.5$ reconstruction has significantly
less power than a scale-free H-Z spectrum on scales $k\ga0.1\,\impc$,
while the $\Gamma=0.2$ reconstruction is closer to a H-Z spectrum but
has relatively more power. However as we shall see the latter
possibility does not give a good fit to the Boomerang/MAXIMA data with
the BBN value (\ref{omegab}) of the baryon density. To ensure
compatibility with BBN {\em requires} a break in the primordial
spectrum.

\subsection{Acoustic peaks}

In linear theory the co-efficents $C_l$ in the spherical harmonics
expansion of the CMB anisotropy are a linear projection of $P(k)$, so
the effect of changing $P^0(k)$ would just be to proportionally alter
the angular power for the multipoles $l$ corresponding to the relevant
values of $k$. We can approximate $l\simeq\pi/\theta_\LSS$ where
$\theta_\LSS\simeq\lambda/D_\LSS$ is the angular size corresponding to
a comoving distance $\lambda$ at the last scattering surface (LSS),
which is at a distance $D_\LSS\simeq2(c/H_0\Omega_\rmm^{0.4})$ for a
flat universe \cite{vs92}. Thus
\begin{equation} 
 l \simeq k/D_\LSS \simeq 3000 k \Omega_\rmm^{-0.4} h^{-1}. 
\end{equation}
For the $\Gamma=0.5$ case (corresponding e.g. to $\Omega_\rmm=0.805$,
$\Omega_\Lambda=0.195$) the inferred primordial power in
Figure~\ref{fig1} is suppressed by a factor $\ga2$ for $k\ga0.1\,\impc$,
implying a similar decrease in the CMB anisotropy at $l\ga500$. For
the $\Gamma=0.2$ case (corresponding e.g. to $\Omega_\rmm=0.35$,
$\Omega_\Lambda=0.65$), the inferred primordial spectrum is a factor
of $\sim2$ above the H-Z one at $k\simeq0.04\,\impc$ so the CMB
anisotropy should be proportionally boosted at $l\ga250$.

\begin{figure}
\epsfxsize\hsize\epsffile{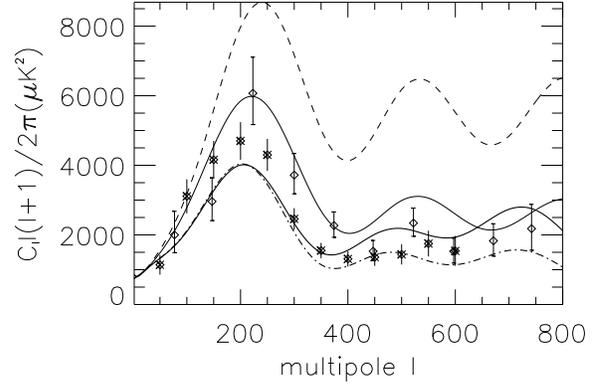}
\caption[junk]{The predicted CMB angular power spectra corresponding
 to the reconstructed primordial spectra for $\Gamma=0.5$ (dot-dashed
 line) and $\Gamma=0.2$ (dashed line). The corresponding results for a
 primordial H-Z spectrum is also shown for $\Gamma=0.5$ and
 $\Gamma=0.2$ (thick and thin full lines). The data are from the
 Boomerang (crosses) and MAXIMA (diamonds) experiments.}
\label{fig2}
\end{figure}

To check these expectations, we ran the Boltzmann code CMBFAST
\cite{sz96} with the above reconstructed primordial spectra as well as
a H-Z spectrum with the same cosmological parameters. As seen in
Figure \ref{fig2}, the angular power resulting from the APM inversion
with $\Gamma=0.5$ is indeed depressed below the H-Z case from the
second acoustic peak onwards, as noted by Adams et al. (1997b). By
contrast, for the $\Gamma=0.2$ reconstruction all the acoustic peaks
are boosted above the H-Z case. It is clear that the recent CMB
observations favour a high density universe. Moreover the observed
suppression of the second acoustic peak in the Boomerang/MAXIMA data
confirms the prior expectation for a primordial density perturbation
with broken scale-invariance. To quantify this we now perform detailed
fits to the data. 


\subsection{Fits to the CMB and LSS data} 

We parameterise the `step' in the primordial power spectrum (see
Figure~\ref{fig1}) as:
\begin{eqnarray}
 P^0(k)= \left\{ \begin{array}{ll}
         A k,          & k \leq k_\sta \\
         C k^{\alpha}, & k_\en \leq k \leq k_\sta \\
         B k,          & k \geq k_\en \\
                \end{array}\right.
\label{pkmodel}
\end{eqnarray}
where $k_\sta$ and $k_\en$ mark the start and end of the break from a
H-Z spectrum, with amplitudes $A$ and $B$. (The values of $C$ and
$\alpha$ are specified by the other parameters.) In the multiple
inflation model (Adams et al. 1997b), the actual form of the spectrum
during the phase transition is difficult to calculate since the usual
`slow-roll' conditions are violated. However a robust expectation is
that
\begin{equation}
 0.5 \la \ln \left(\frac{k_\en}{k_\sta}\right) \la 2,
\label{dur}
\end{equation}
because the field undergoing the symmetry-breaking phase transition
evolves exponentially fast to its minimum. The ratio of the amplitudes
$A/B$ is determined by the (unknown) superpotential couplings of the
field undergoing the phase transition but is expected to exceed unity
(i.e. there is a decrease in the power).

We allow the spectral and cosmological parameters to range rather
widely as follows:

\begin{itemize}

\item $k_\sta$ from (0.01--0.15) $\impc$ 

\item $\ln(k_\en/k_\sta)$ from 0.01--4 

\item $A/B$ from 0.3--7.2 

\item $\Omega_\Lambda$ from 0--0.7 (keeping $\Omega_\rmm+\Omega_\Lambda=1$)

\item $h$ from 0.45--0.9 

\item $\Omega_\rmB h^2$ from 0.0014--0.033 

\end{itemize}
We also consider a bias parameter $b$, defined as the square root of
the ratio of the APM and CMB normalisations (which is expected to be
close to unity, see Appendix~A).

For each combination of the above parameters we construct the
primordial power spectrum and then obtain the matter power spectrum by
convoluting with the CDM transfer function calculated by CMBFAST
(which differs slightly from the analytic approximation in
Equation~\ref{Tk}). It is also used to calculate the expected CMB
angular power spectrum. These are then compared, respectively, with
the APM data \cite{gb98},\footnote{We use the values of $P_\APM(k)$
given in Table~2 with errors estimated from the variance in the 4
disjoint sky regions in the catalogue. Ideally we should use the
linear spectrum $P_{\lin}(k)$ recovered for each specific choice of
cosmological parameters. However the two are very similar since we
restrict ourselves to the quasi-linear range $k\la0.2\,\impc$ (see Fig
\ref{nlbias} in Appendix~A), and the estimated errors in $P_\APM(k)$
in this range are large enough to encompass all possible linear
spectra.} and the decorrelated COBE \cite{th97} and Boomerang data
\cite{boom00} \footnote{There appear to be systematic differences
between the Boomerang and MAXIMA data as shown in Figure~\ref{fig2}.
However according to Jaffe et al (2000), the two datasets are quite
consistent within their respective calibration uncertainties (20\% and
8\%, $1\sigma$ in $C_l$) and beam uncertainties (10\% and 5\%). For
definiteness we use the Boomerang data alone; clearly our results
would be very similar if we use the MAXIMA data instead.} using
appropriate window functions, to determine
$\chi^2(\alpha)=\sum_{i}[({\rm model}(\alpha)-{\rm
data}(i))/\sigma(i)]^2$. We use 19 CMB data points [7 from COBE
(excluding the anomalously low quadrupole) and 12 from Boomerang] and
10 APM data points (for $0.01\,\impc\la{k}\la0.2\,\impc$). We quantify the
goodness-of-fit following the procedure of Lineweaver \& Barbosa
(1998).

In Table~\ref{tab1} we show the values of the Hubble constant, baryon
density, and amplitude change at the spectral break which minimise the
$\chi^2$ (for a specific choice of $\Omega_\Lambda$) for fits to the
CMB data {\em alone}. The step in the primordial spectrum is seen to
be not essential for a good fit, in particular even with $A/B=1$
(i.e. no step) the $\chi^2$ does not worsen significantly in most
cases. The best overall fit obtains for $\Omega_\Lambda=0.5$,
consistent with the analysis by the Boomerang collaboration
\cite{boom00}. Note however that the baryon density required is 65\%
higher than the preferred BBN value (\ref{omegab}) although still
within the upper limit (\ref{omegabmax}) set by interstellar
deuterium.

\begin{table}
\begin{tabular}{|c||c|c|c|c|c|c|} \hline
$\Omega_\Lambda$ & $h$ & $k_\sta$ & $\ln\left(\ffrac{k_\en}{k_\sta}\right)$ 
 & $\Omega_{\rmB}h^2$ & $\ffrac{A}{B}$ 
 & $\chi^2$ $\left(\ffrac{A}{B}=1\right)$ \\ \hline
\hline
0.0 & 0.45 & 0.03 & 0.7 & 0.019 & $\ge$ 1.7 & 7.5 (30)\\ \hline
0.2 & 0.55 & 0.03 & 0.7 & 0.023 & $\ge$ 1.4 & 6.9 (19)\\ \hline
0.3 & 0.75 & 0.05 & 0.8 & 0.031 & $\ge$ 1.3 & 6.6 (14)\\ \hline
0.4 & 0.75 & 0.03 & 0.8 & 0.031 & $\ge$ 1.1 & 6.3 ( 8)\\ \hline
0.5 & 0.80 & 0.02 & 0.6 & 0.031 & $\ge$ 1.0 & 5.8 ( 6)\\ \hline
0.6 & 0.85 & 0.02 & 0.7 & 0.031 & $\ge$ 1.0 & 5.5 ( 6)\\ \hline
0.7 & 0.90 & 0.03 & 1.0 & 0.031 &  0.7--0.9 & 7.5 (11)\\
\hline
\end{tabular}
\caption[junk]{Parameters for best fits to CMB data (13~$\dof$).}
\label{tab1}
\end{table}

\begin{table}
\begin{tabular}{|c||c|c|c|c|c|} \hline
$\Omega_\Lambda$ & $h$ & $k_\sta$ & $\ln\left(\ffrac{k_\en}{k_\sta}\right)$ 
 & $\ffrac{A}{B}$ &
$\chi^2$ \\ \hline
\hline
0.0  &  0.50  &  0.04  &  0.7  &  1.7-3.1 & 8.1  \\ \hline
0.2  &  0.50  &  0.03  &  0.7  &  1.7-3.1 & 7.2  \\ \hline
0.3  &  0.55  &  0.03  &  0.8  &  1.7-3.1 & 6.9  \\ \hline
0.4  &  0.60  &  0.03  &  0.8  &  1.7-3.1 & 6.9  \\ \hline
0.5  &  0.65  &  0.04  &  0.6  &  1.7-2.8 & 7.9  \\ \hline
0.6  &  0.75  &  0.05  &  0.6  &  1.5-2.1 & 11.9 \\ \hline
0.7  &  0.80  &  0.01  &  0.6  &  0.7-0.9 & 35.1 \\ \hline
\end{tabular}
\caption[junk]{Parameters for best fits to CMB data with the
 BBN constraint on the baryon density.}
\label{tab2}
\end{table}

\begin{figure}
\epsfxsize\hsize\epsffile{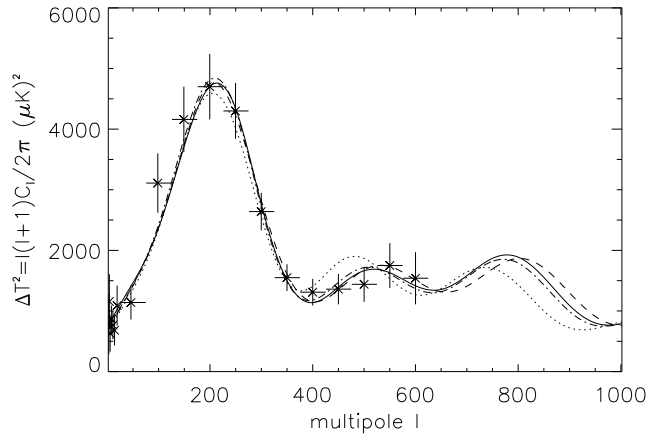}
\caption[junk]{CMB angular power spectra for the best-fit models in
 Table~\protect\ref{tab2}, with $\Omega_\Lambda=0$ (continuous line),
 $\Omega_\Lambda=0.2$ (dashed line), $\Omega_\Lambda=0.4$ (dot-dashed
 line), and $\Omega_\Lambda=0.6$ (dotted line). The data shown are
 from COBE and Boomerang. All models have the BBN baryon density
 $\Omega_{\rmB}=0.019h^{-2}$.}
\label{fig3}
\end{figure}

If we now demand that the models should have the BBN baryon density
(\ref{omegab}), the Hubble parameter in the range (\ref{h}) and a
spectral break duration (\ref{dur}), then a step in the primordial
spectrum is definitely indicated as seen from Table~\ref{tab2}. [In
particular setting $A/B=1$ (no step) always gives an unacceptably high
value of $\chi^2>26$.] The angular power spectra corresponding to some
of these models is shown in Figure~\ref{fig3} along with the Boomerang
and COBE data. We emphasize that a good fit is now obtained even for
$\Omega_\Lambda=0$.

\begin{table}
\begin{tabular}{|c||c|c|c|c|c|c|c|} \hline
$\Omega_\Lambda$ & $h$ & $\Omega_{\rmB}h^2$ & $k_\sta$ 
& $\ln\left(\ffrac{k_\en}{k_\sta}\right)$ & $\ffrac{A}{B}$ & $\chi^2$ \\ \hline
\hline
0.0  &  0.50  &  0.019  &  0.07   &  1.5  &  4.3-6.5  & 11.9\\ \hline
0.2  &  0.60  &  0.023  &  0.07   &  1.4  &  3.7-5.6  &  9.4\\ \hline
0.3  &  0.65  &  0.023  &  0.07   &  1.4  &  3.6-5.5  &  8.2\\ \hline
0.4  &  0.70  &  0.023  &  0.07   &  1.4  &  3.2-5.0  &  8.3\\ \hline
0.5  &  0.85  &  0.031  &  0.08   &  1.2  &  3.1-4.6  &  7.5\\ \hline
0.6  &  0.90  &  0.031  &  0.07   &  1.3  &  2.4-3.6  &  9.0\\ \hline
0.7  &  0.90  &  0.023  &  0.06   &  2.0  &  1.6-5.1  & 30.4\\ \hline
\end{tabular}
\caption[junk]{Parameters for best fits to CMB+APM data (22~$\dof$).}
\label{tab3}
\end{table}

Next we fit simultaneously to both the APM and CMB data allowing for a
small difference in the normalisations, i.e. a bias $b$. We have taken
into account that the mean redshift of the APM galaxies is $z=0.15$ so
a corresponding correction for the growth factor to $z=0$ needs to be
applied. It is seen from Table~\ref{tab3} that the best overall fit is
still obtained for $\Omega_\Lambda=0.5$, although other values remain
quite acceptable. In particular the $\Omega_\Lambda=0$ model requires
a Hubble parameter at the low end of the allowed range but has the
advantage of being consistent with the BBN value of the baryon
density. We emphasise that a break in the spectrum is now definitely
required since the value of $\chi^2$ exceeds 45 otherwise. The most
likely duration of the break also comes out in accord with the
theoretical expectation (\ref{dur}).

\begin{table}
\begin{tabular}{|c||c|c|c|c|c|c|} \hline
$\Omega_\Lambda$ & $h$ & $k_\sta$ & $\ln\left(\ffrac{k_\en}{k_\sta}\right)$ 
 & $\ffrac{A}{B}$ & $\sigma_8$ & $\chi^2$ \\ \hline
\hline
0.0  &  0.50  &  0.07   &  1.5  &  4.3-6.5  &  0.65-0.73  &
  11.9\\ \hline
0.2  &  0.55  &  0.06   &  1.6  &  3.6-5.6  &  0.69-0.78  &
  10.0\\ \hline
0.3  &  0.60  &  0.06   &  1.6  &  3.4-5.0  &  0.76-0.85  &
 10.3\\ \hline
0.4  &  0.65  &  0.05   &  1.5  &  3.1-4.6  &  0.77-0.87  &
 13.1\\ \hline
0.5  &  0.70  &  0.05   &  1.6  &  2.6-4.1  &  0.81-0.93  &
 18.8\\ \hline
0.6  &  0.75  &  0.05   &  2.0  &  2.0-2.8  &  0.90-0.99  &
 29.7\\ \hline
\end{tabular}
\caption[junk]{Parameters for best fits to CMB+APM data
 with the BBN constraint on the baryon density.}
\label{tab4}
\end{table}

\begin{figure}
\epsfxsize\hsize\epsffile{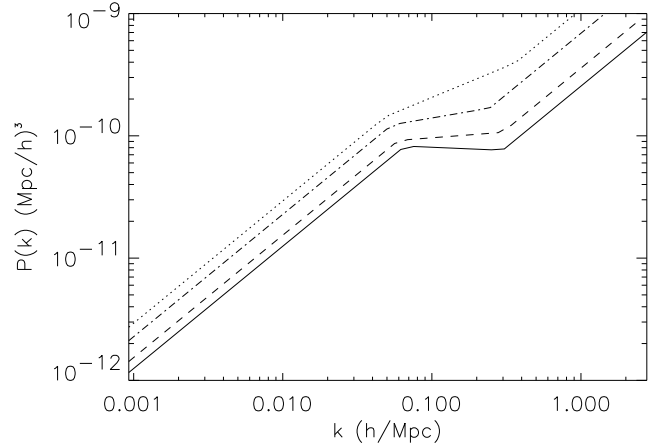}
\caption[junk]{Primordial power spectra for the best-fit models in
 Table~\ref{tab4} with $\Omega_\Lambda=0$ (continuous line),
 $\Omega_\Lambda=0.2$ (dashed line), $\Omega_\Lambda=0.4$
 (dot-dashed line), and $\Omega_\Lambda=0.6$ (dotted line). All
 models have the BBN baryon density $\Omega_{\rmB}=0.019h^{-2}$.}
\label{fig4}
\end{figure}

\begin{figure}
\epsfxsize\hsize\epsffile{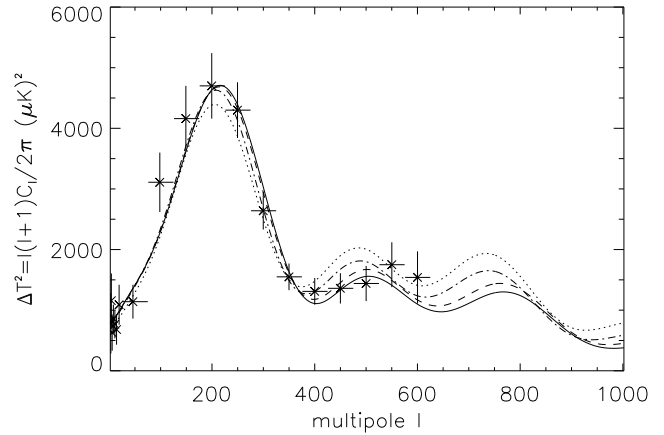}
\caption[junk]{CMB angular power spectra for the best-fit models in
 Table~\ref{tab4} with $\Omega_\Lambda=0$ (continuous line),
 $\Omega_\Lambda=0.2$ (dashed line), $\Omega_\Lambda=0.4$
 (dot-dashed line), and $\Omega_\Lambda=0.6$ (dotted line). All
 models have the BBN baryon density $\Omega_{\rmB}=0.019h^{-2}$. The
 data points are from COBE and Boomerang.}
\label{fig5}
\end{figure}

\begin{figure}
\epsfxsize\hsize\epsffile{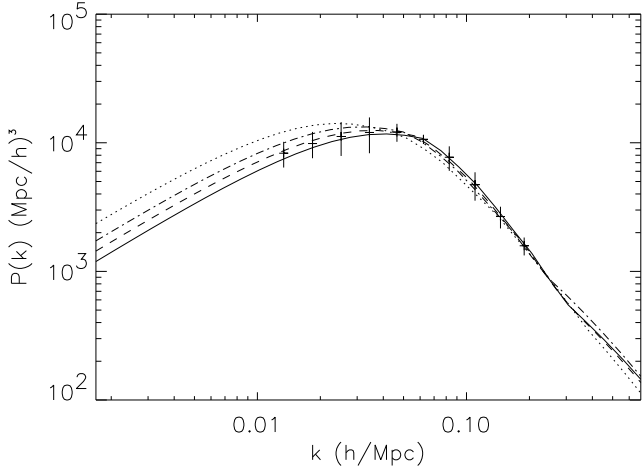}
\caption[junk]{Matter power spectrum for the best-fit models in
 Table~\ref{tab4} with $\Omega_\Lambda=0$ (continuous line),
 $\Omega_\Lambda=0.2$ (dashed line), $\Omega_\Lambda=0.4$ (dot-
 dashed line), and $\Omega_\Lambda=0.6$ (dotted line). All models
 have the BBN baryon density $\Omega_{\rmB}=0.019h^{-2}$. The data are
 from the APM survey.}
\label{fig6}
\end{figure}

\begin{figure*}
\epsfxsize\hsize\epsffile{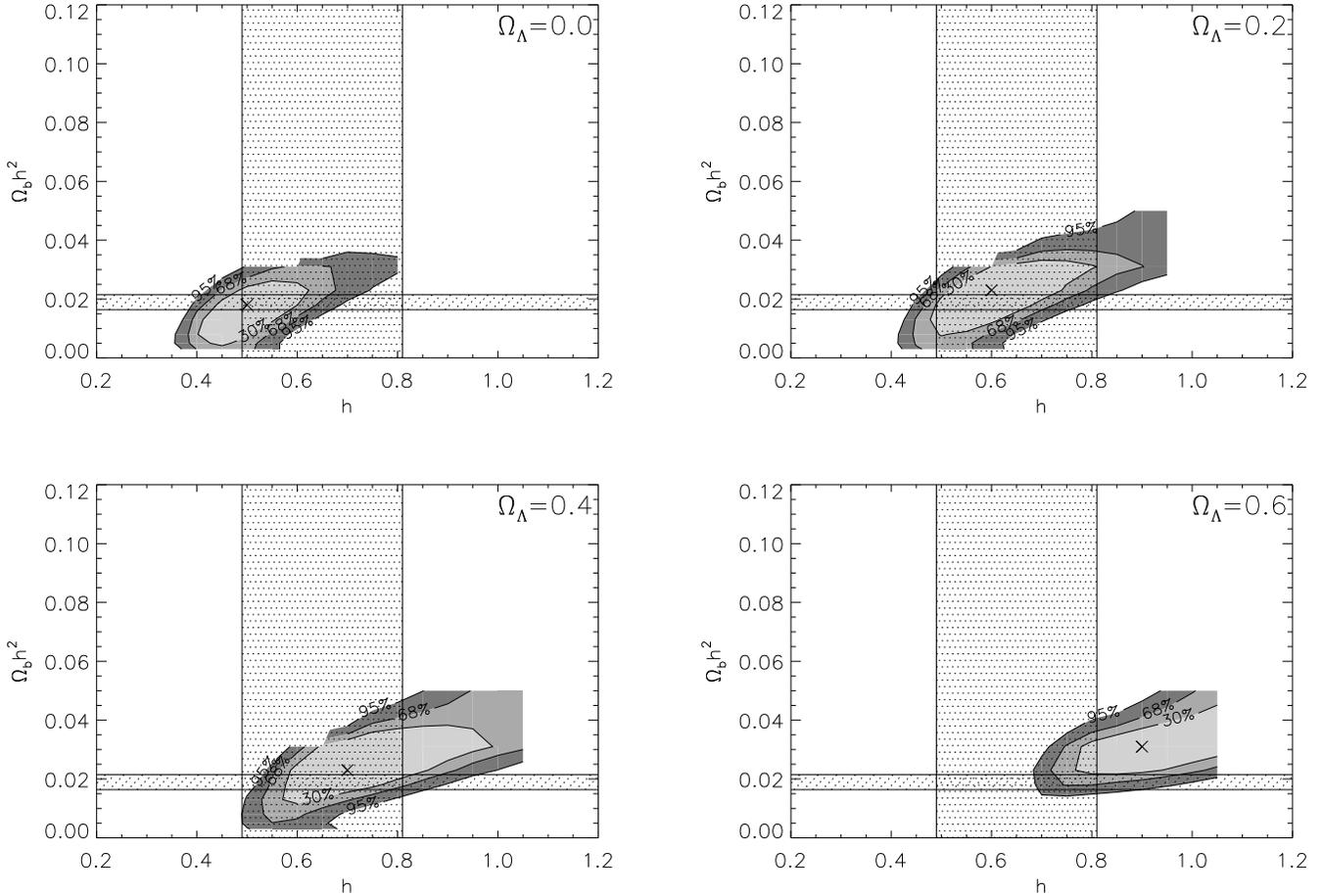}
\caption[junk]{Goodness-of-fit contours for the joint fit to the CMB
 and APM data of a primordial spectrum with a `step', for different
 choices of $\Omega_\Lambda$. The vertical and horizontal shaded bands
 indicate the constraints (\protect\ref{h}) and (\protect\ref{omegab})
 on $h$ and $\Omega_{\rmB}h^2$ respectively while the crosses indicate
 the best fits from Table~\protect\ref{tab3}. Values of
 $\Omega_\Lambda\ga0.6$ provide an unacceptably poor fit to the
 data.}
\label{fig7}
\end{figure*}

Finally Table~\ref{tab4} shows the result of imposing the BBN
constraint (\ref{omegab}), and the (less important) constraints on the
Hubble parameter (\ref{h}) and the duration (\ref{dur}). We have also
required that the bias $b$ be within 20\% of unity (see Appendix~A). We
see that the data now prefer lower values of $\Omega_\Lambda$ (and
$h$). The magnitude of the required spectral break decreases with
increasing $\Omega_\Lambda$ (Figure~\ref{fig4}) but one cannot do
without such a break altogether since otherwise the $\chi^2$ exceeds
53. Figures~\ref{fig5} and \ref{fig6} show, respectively, the fits to
the CMB and APM data for these models (using the mean value of the
break amplitude $A/B$). In Figure~\ref{fig7} we display the
goodness-of-fit contours in the $\Omega_{\rmB}h^2-h$ plane for the
joint fit to the CMB and APM data; the contours are ragged because we
have not used a fine enough grid of parameter values.  We see that
agreement with the BBN value (\ref{omegab}) of the baryon density can
be achieved only for low $\Omega_\Lambda$. In fact a critical density
matter-dominated universe is favoured with the Hubble parameter at its
lower limit (so that the age of the universe is acceptable at
$\sim13$~Gyr).

\subsection{Implications for cluster abundances}

In Table~\ref{tab4} we show also the value of the variance $\sigma(R)$
in (dark matter) fluctuations (normalised to the CMB), over a sphere
of size $R=8\,\mpc$:
\begin{equation}
 \sigma^2 (R) 
 = \frac{1}{H_0^4} \int_0^\infty 
   W^2(kR) ~\delta_\H^2(k)~ T^2(k)~ k^3 \rmd k, 
\label{sigma8}
\end{equation}
using a `top hat' smoothing function,
$W(kR)=3[\ffrac{\sin(kR)}{(kR)^3}-\ffrac{\cos(kR)}{(kR)^2}]$. (The
range of $\sigma_8$ for a given value of $\Omega_\Lambda$ corresponds
to the range of the step-size $A/B$.) The tabulated values are
systematically somewhat higher than the mean value
$\sigma_8=0.56\Omega_\rmm^{-0.47}$ for flat cosmologies inferred from
the observed abundances of X-ray emitting clusters, although within
the 95\% c.l. upper limit which is higher by
$20\Omega_\rmm^{0.2\log\Omega_\rmm}$ per cent \cite{vl99}. We wish to
emphasise that the break in the primordial spectrum implies a much
lower $\sigma_8$ than the corresponding H-Z spectrum with the {\em
same} cosmological parameters. For example the COBE-normalised sCDM
(H-Z) model with $\Omega_m\simeq1$ gives $\sigma_8\simeq1.4$
(e.g. Stompor, Gorski and Banday 1996) as opposed to $\sigma_8\simeq
0.65-0.73$ in Table~\ref{tab4}, which is much closer to the value of
$\sigma_8\simeq0.56$ inferred from the cluster abundance. As shown in
Figure~\ref{fig4}, this is because of the decrease in power at small
scales relative to a scale-invariant spectrum. Note that we cannot
constrain well the region $k>0.2\,\impc$ (corresponding to $R<8\,\mpc$),
because of uncertainties in the reconstruction of the primordial
spectrum due to non-linear biasing and non-linear gravity effects. For
simplicity we have assumed a H-Z spectrum for $k>0.6\,\impc$
(Equation~\ref{pkprim}) so in principle other (smaller) values of
$\sigma_8$ are also allowed. For instance, if there were further
breaks in the primordial spectrum, as is quite possible, then a better
agreement could easily be achieved. In order to constrain such
possibilities we intend to investigate in further work the
Lyman-$\alpha$ forest $P(k)$ and its evolution with redshift
\cite{croft99} which provides a probe of such small scales in the {\em
linear} regime.
                                                 
\begin{figure}
\epsfxsize\hsize\epsffile{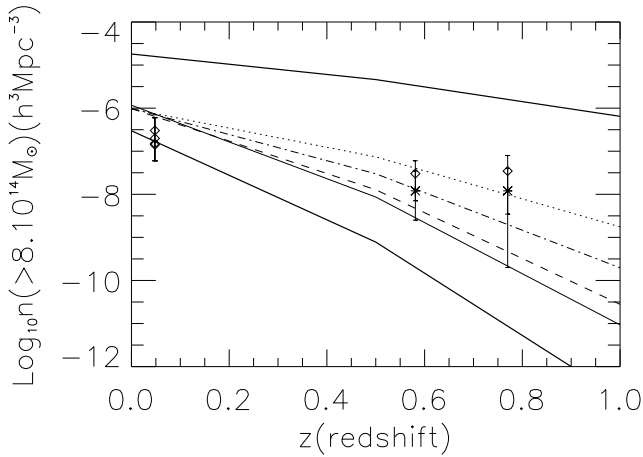}
\caption[junk]{The number density of rich clusters with $M>8\times
 10^{14}M_\odot$ as a function of redshift. Points with errorbars
 correspond to the observations as depicted in Bahcall \& Fan
 (1998). The thick full lines correspond to the P-S prediction for the
 spatially flat, COBE-normalised H-Z model with $\Omega_\Lambda=0.2$,
 $\Omega_m\simeq0.8$ (top line), and scaled to match the cluster
 abundance at $z=0$ (bottom line). The curves in between correspond to
 the best-fit models in Table~\ref{tab4} with $\Omega_\Lambda=0$ (full
 line), $\Omega_\Lambda=0.2$ (dashed line), $\Omega_\Lambda=0.4$
 (dot-dashed line), and $\Omega_\Lambda=0.6$ (dotted line).}
\label{fig8}
\end{figure}

In Figure~\ref{fig8} we show the predictions for the redshift
evolution of the cluster abundance in our models according to the
Press-Schechter (P-S) theory \cite{ps74}, compared to the observations
as presented in Bahcall \& Fan (1998). We consider clusters with mass
$M>8\times10^{14}M_{\odot}$ (neglecting the $R<R_{\rm com}=1.5\,\mpc$
condition). The spatially flat COBE-normalised H-Z model with
$\Omega_\Lambda=0.2$, $\Omega_m\simeq0.8$ (top thick full line) does
not match the observations, a fact that has been used to rule out this
model. Even when this model is normalised at $z=0$ by lowering its
amplitude to $\sigma_8\simeq0.6$ (bottom thick full line), it still
fails to reproduce the observed redshift evolution in the cluster
abundances \cite{bf98}. By contrast, the best-fit models in
Table~\ref{tab4} do match the redshift evolution, especially if we
allow for the possibility mentioned above of lowering the $\sigma_8$
values further to better match the $z=0$ data.\footnote{Note that the
P-S prediction depends not only on the value of $\sigma_8$ at $z=0$
and the corresponding cosmological evolution of the variance, but also
on the shape of the spectrum over the mass scales considered. In our
case, these scales correspond to $R>10\,\mpc$, i.e. near the break in
the primordial spectrum.} We emphasise that these predictions have not
been adjusted in any way to fit cluster abundances; all parameters
have been fixed already by the fit to the APM and CMB data. It is
interesting to note that even the $\Omega_\Lambda=0$ model can now
reproduce the slow growth of the cluster abundance with redshift due
to the break in the primordial spectrum.

\section{Discussion}

Several authors have in the past noted the possibility of generating
spectral features in the H-Z density perturbation by fine tuning the
parameters of the inflaton potential \cite{kl87,sbb89,hbkp90,hb90}. We
would like to emphasise that in contrast to such ``designer'' models,
the break in the H-Z spectrum discussed here is generated by a {\em
physical} mechanism, viz. supersymmetry breaking during inflation
(Adams et al. 1997b, Ross 1998). As noted subsequently
\cite{lp97,kksy00}, the resulting damping of the CMB anisotropy is
also a characteristic of `double inflation' which occurs in toy models
wherein inflation is driven in two stages --- first by higher-order
gravity and then by a scalar field \cite{kls85,kp88,gms91}, or by two
coupled scalar fields \cite{s85,kl87,ps92}. The model parameters can
be tuned to generate a break in the primordial spectrum at any chosen
scale. The advantages of this for fitting LSS observations was first
emphasised by Silk \& Turner (1987) and have been explored thoroughly
\cite{ps92,gms94,p94,pps94,agmm95,ps95,sm96}, although these authors
did not discuss the implications for the CMB. A spectral feature
similar to that observed can also be generated in a toy model where
the inflaton evolves through a kink in its potential \cite{s92},
leading to similar suppression of the secondary acoustic peaks
\cite{aegms97,lps98,lpp99}. Recently there have been efforts to
progress beyond toy models and implement multiple inflation in
supersymmetric models \cite{st98,l98,l00}. Spectral features can also
be generated by resonant production of particles during inflation
\cite{ckrt00}.

Silk \& Gawiser (2000) have noted independently that it is difficult
for a $\Lambda$CDM model to fit LSS and CMB observations
simultaneously without considering departures from scale-invariance;
they advocate adding a `bump' at $k\sim0.06\,\impc$ to a primordial
H-Z spectrum.\footnote{Griffiths, Silk \& Zaroubi (2000) have very
recently shown that this can also fit the recent CMB and LSS data;
however they choose to discount the APM data in favour of the (less
precise) power spectrum inferred from the PSCz survey \cite{ht00}
which does not show a `shoulder' at $k\sim0.1\,\impc$. We would point
out that our models in Table~\ref{tab4} also fit the PSCz power
spectrum.}  Einasto et al. (1999) have also concluded from an
exhaustive analysis of LSS data that a scale-free primordial spectrum
is excluded. In particular it has been noted that the power spectrum
of clusters is better fitted if the primordial spectrum has a
step-like feature at $k\sim0.1\,\impc$ \cite{gs99}. The implications
of broken scale-invariance for accurate determination of cosmological
parameters using forthcoming LSS and CMB data have been investigated
\cite{wss99,h00}.

It is clear that the notion of a scale imprinted on the primordial
density perturbation is favoured by a number of observational
considerations as well as having a natural physical interpretation in
the context of realistic inflationary models. In this paper we have
demonstrated that the recent small angular-scale CMB anisotropy data
favour a deviation from scale-invariance in the range
$k\sim(0.06-0.6)\impc$ {\em independently} of the similar previous
indication from studies of LSS, in particular the APM survey. This is
reassuring since the latter rests on the assumption (for which
arguments are given in Appendix~A) that there is no scale-dependent
bias between APM galaxies and dark matter.
 
Allowing the primordial spectrum to depart from scale-invariance has
important consequences for the deductions that can be made
about cosmological parameters from LSS and CMB data. The standard
interpretation of LSS data (e.g. APM $P(k)$ and $\sigma_8$ from
cluster abundances) favours a low $\Gamma\sim\Omega_m{h}\sim0.2$
universe assuming a COBE-normalised H-Z spectrum. However as we have
shown these observations can be satisfactorily accounted for with much
larger values of $\Omega_m$ if we allow for a break in the primordial
spectrum. Our best fit models in Table~\ref{tab4}, which obey the BBN
constraint (\ref{omegab}) on the baryon density, prefer higher values
of $\Omega_m$ (and lower values of $\Omega_\Lambda$). The required
amplitude of the break decreases with increasing $\Omega_\Lambda$ but one
cannot do without such a break altogether. In particular the value
$\Omega_\Lambda\sim0.7$ favoured by the SN~Ia data \cite{scp99,hzs98}
is {\em not} permitted, while it is quite possible to have a universe
with zero cosmological constant and $\Omega_m=1$. Of course the
high observed fraction of baryons in clusters, combined with the BBN
value of the baryon density (\ref{omegab}), implies an independent
limit on the matter density \cite{wnfw93}; a recent analysis quotes
$\Omega_\rmm<0.31\,h^{-1/2}$ at 95\% c.l. \cite{mhh00}. There are many
assumptions made in such analyses, viz. that clusters are spherical
and in hydrostatic equilibrium with an isothermal temperature profile,
that the measured baryon fraction reflects the universal value, that
there is no preheating before or energy injection after cluster
formation, etc. It is important to reassess these issues in the light
of our increasing understanding of galaxy and cluster formation to
quantify the systematic uncertainties better. 

The hypothesis of a primordial density perturbation with broken
scale-invariance, although radical, is eminently falsifiable. The
ongoing 2DF and SDSS redshift surveys can confirm or rule out such a
feature in the power spectrum of galaxy clustering, while the
forthcoming MAP and Planck missions will determine whether {\em all}
the secondary acoustic peaks in the CMB angular spectrum are indeed
suppressed as expected. By contrast the alternative hypothesis of a
baryon density $\sim65\%$ higher than the BBN value predicts a {\em
boosted} third peak (e.g. Tegmark \& Zaldariagga 2000), so this would
be a definitive test distinguishing between the two possibilities. We
note that in either case there are important implications for the
physics of the early universe. To increase the baryon-to-photon ratio
after BBN by $\sim65\%$ requires a {\em decrease} in the comoving
entropy; the only mechanism suggested which can achieve this proposes
that the photons in our universe become cooled through their coupling
to a `shadow' universe \cite{bh91}. However this possibility, recently
invoked by Kaplinghat and Turner (2000), was shown to be severely
constrained, if not ruled out altogether, from considerations of the
concommitant spectral distortion in the CMB together with the bound
from Supernova 1987a on new weakly interacting particles
\cite{bs97}. By contrast broken scale-invariance has a natural
explanation in a phase transition occuring during inflation as
expected in supersymmetric theories. If established this would provide
the first direct connection between astronomical data and physics at
very high energies.

\section*{Acknowledgments}
We thank Pedro Ferreira, Joe Silk, Graham Ross and Rom\'an Scoccimarro
for stimulating discussions. J.B. and E.G. acknowledge grants from
IEEC/CSIC and DGES(MEC) (Spain) --- project PB96-0925 and Acci\'on
Especial ESP1998-1803-E; J.B. would also like to thank INAOE for their
warm hospitality. The work of M.G.S. was supported by the
Funda\c{c}\~{a}o para a Ciencia e a Tecnologia (Portugal) under
program PRAXIS XXI/BD/18305/98.

\newpage
\appendix

\section{The effects of biasing}\label{sec:bias}

Independently of how or where galaxies form they must eventually fall
into the dominant gravitational wells and thus trace out the
underlying mass distribution (see Peebles 1980, Fry 1996). However, in
principle, the galaxy distribution may be `biased' with respect to the
underlying mass fluctuations (e.g. Bardeen et al. 1986). There are
several observations indicating that this effect is in fact small. The
theoretical predictions for the first few connected moments, based on
this hypothesis \cite{jbc93,b94a,b94b} are in good agreement with the
APM data \cite{g94,gf94,fg99}. The current precision of this higher
order correlation test is 20\% and expected to improve with further
data. The APM angular correlation function $\xi_g(r)$ shows an
inflection point near $\xi_g\simeq1$ very similar to the long awaited
`shoulder' in $\xi\simeq1$, generated by gravitational dynamics
\cite{gr75}. The agreement between the two characteristic scales can
be used to constrain the linear biasing factor for the APM catalogue
to be within 20\% of unity \cite{gj00}. We summarise these arguments
\cite{gb98} below for completeness since the assumption of small or no
bias is crucial for the present study.

Let us assume that the (smoothed) galaxy fluctuations $\delta_g$ are
related to those in the mass $\delta_m$ by a local transformation,
$\delta_g(x)= F[\delta_m(x)]$, which can be expanded as a Taylor
series: $F = b_1 \delta_m + b_2 \delta_m^2+\ldots$. Then the 2-point
function $\xi_2^g(r)\equiv\langle\delta_g(x)\delta_g(x+r)\rangle$ on
the scale $r$ will be:
\begin{eqnarray}
 \xi_2^g(r) 
 &=& b_1^2 \xi_2^m (r) + b_1 b_2 \langle\delta_m(x)\delta_m(x+r)^2\rangle
 \nonumber \\
 &~& + b_1 b_2 \langle\delta_m^2(x)\delta_m(x+r)\rangle + \ldots ,
\end{eqnarray}
where all further terms are of order 4 or higher in $\delta_m$ and
correspond therefore to either higher order correlations, $\xi_J$ with
$J>2$, or higher powers in $\xi_2$. If $\delta_m$ is Gaussian or
hierarchical (as is the case for evolution through gravity), the
higher order correlations $\xi_J$ are at most of order
$\xi_2^{J-1}$. This means that at large scales where $\xi_2<1$, the
first term is dominant so that only the amplitude of the 2-point
statistics, but {\em not} its shape, may be altered by biasing. This
has been confirmed in N-body simulations and galaxy biasing models
\cite{mjw97,sshj00}.

\begin{figure}
\epsfxsize\hsize\epsffile{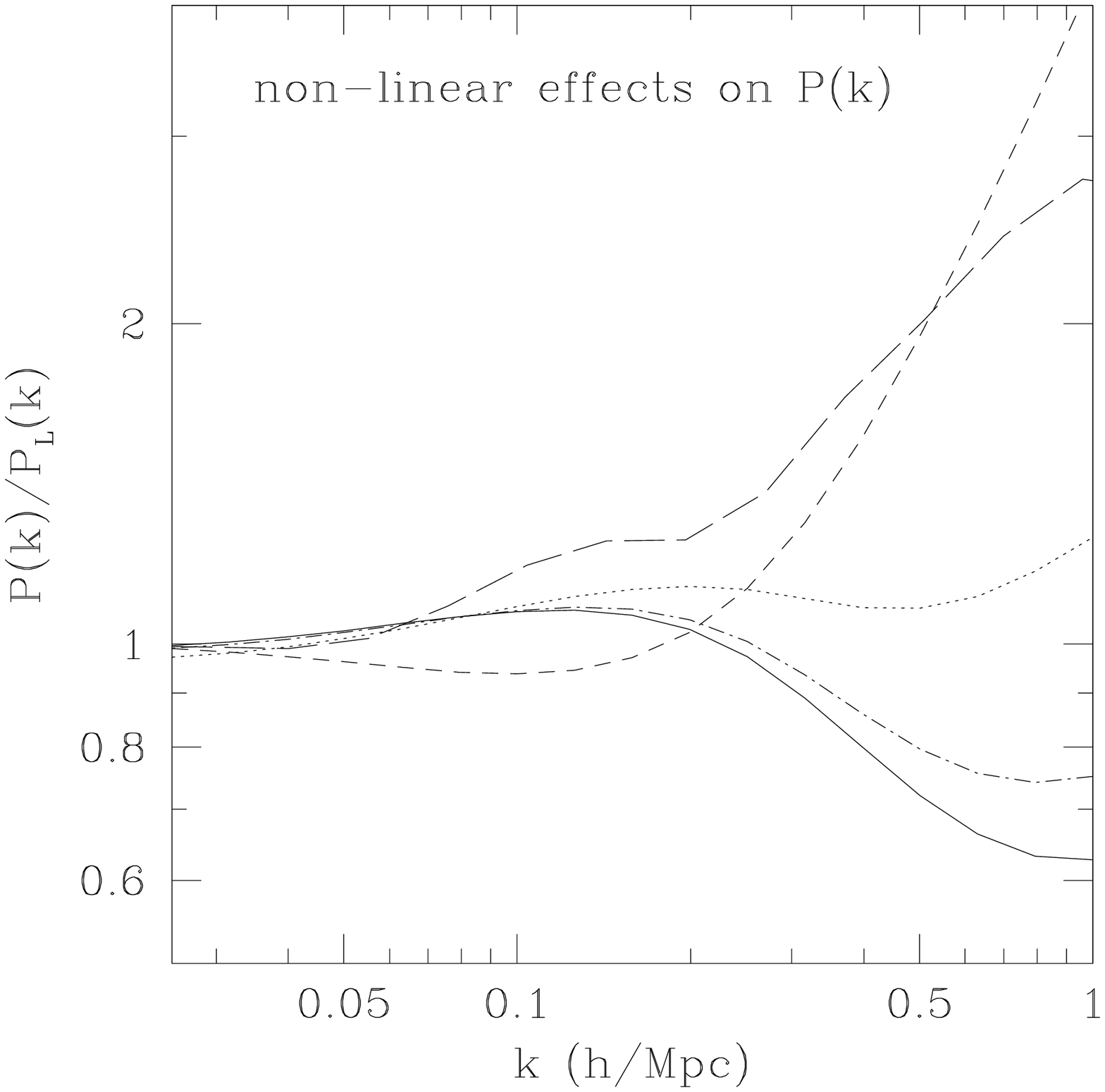}
\caption[junk] {The ratio of the non-linear to linear power spectrum
 illustrating non-linear effects on $P(k)$. Long and short-dashed
 lines show non-linear effects in the mass while the other lines show
 non-linear effects in the final galaxy distribution for various halo
 models (from Scoccimarro et al 2000).}
\label{nlbias}
\end{figure}

Thus from the above arguments, the small variance on large scales
($R\ga8\,\mpc$) means that it is reasonable to assume that the shape of
$P(k)$ for galaxies at scales $k\la0.1\,\impc$ coincides with the shape
of the underlying linear matter power spectrum. This argument, which
is based simply on the smallness of the variance and the assumption of
hierarchical clustering, can also be applied to gravity, since the
leading contribution to the correlation functions in perturbation
theory is indeed given by a local transformation (see Fosalba \&
Gazta\~naga 1997). This is clearly illustrated in Figure 11 of
Gazta\~naga and Baugh (1998). By comparing the linear and non-linear
shape of $P(k)$, one can see that it has not been changed
significantly through gravitational evolution on scales where the rms
fluctuations are small, i.e. for $k\la0.1\,\impc$.

How important are non-linearities for $\delta\ga1$, i.e. for
$k\ga0.1\,\impc$? Figure~\ref{nlbias} shows the ratio of the non-linear
to linear power spectrum $P(k)$, normalised to be unity at small $k$
in order to emphasize the non-linear effects. The long-dashed line
corresponds to the APM-like $P(k)$ for $\Omega=1$ dynamics (from
N-body simulations in Baugh \& Gazta\~naga 1996). The short-dashed
line corresponds to a $\Omega=0.3$, $\Lambda=0.7$ $\Lambda$CDM model
(using the non-linear mass fitting formulae). The dotted, continuous
and dot-dashed lines correspond to different non-linear models for
galaxy formation based on halo models (with linear bias $b\la1.5$ for
$\sigma_8\simeq1$) from Figure~6 in Scoccimarro et al (2000). In all
cases non-linear effects become significant ($>10\%$) only at
$k\ga0.2\,\impc$ and are of order unity by $k\sim1\,\impc$. Non-linear
effects seem smaller for galaxies than for the dark matter (in
agreement with the observed $S_J$ in Baugh \& Gazta\~naga 1996). One
could imagine constructing galaxy formation models with stronger
non-linear effects, but it will then be difficult to account for the
observed galaxy higher-order correlations. As pointed out in
Scoccimarro et al (2000), galaxy clustering in the framework of the
halo models is strongly constrained by the measured galaxy
skewness. Thus, given current observational constraints and present
understanding of galaxy formation one can conclude that non-linear
effects cannot have altered the linear $P(k)$ by a factor exceeding
$\sim2$, up to $k\simeq1\,\impc$ .

\end{document}